# A versatile DAQ, monitoring and data processing system for nuclear experiments in CAMAC and VME   standards

Zdeněk Hons

*Acad. Sci. of Czech Republic, Nuclear Physics Institute v.v.i., Řež near Prague, CZ-25068, Czech Republic*
email: hons@ujf.cas.cz

## Abstract

Common and unique features of nuclear physics measurements are examined. Such analysis with respect to existing hardware and software platforms and standards allows to algorithmize  the DAQ, monitoring and processing tasks.

A universal measurement framework is proposed and programmed. It allows to implement various measurement setups via a configuration file where special features are described by a simple language.

An abstract description of the experimental data is introduced and a unified data format is proposed to hide data structures resulting from the various measurement setups. A simple language is developed to describe processing and visualizing the unified data via histograms. A universal processing engine is developed which creates histograms based on requirements described in a configuration file.

# Table of Contents

# 1. Introduction

Experimental research of nuclear and subnuclear phenomena is contituneously evolving process. History of each experiment consists of the design period, hardware assembly, software development and debugging, data taking, data processing, and at last publishing the results. Individual experiments generally do not go one after another, but during the first experiment data processing, the next one is usually being prepared concurrently. Thus the possibility to start from the very beginning occurs very rarely. This makes more difficult to apply approaches offered by e.g. EPICS[6], COMEDI[7] or LabVIEW®[8].

Usually the various measurements have unique proprietary software. Moreover the DAQ and the data processing systems are often designed in such a way which implies their non-trivial adjustment after every modification of the measurement setup. When the author had to manage several such systems simultaneously he found out soon that it can be very time-consuming and tedious process. That's why an attempt has been made to develop a versatile measurement system to avoid the problems mentioned above.

# 2. Abstract description of the DAQ part

In the next paragraphs the measurement process will be analyzed, with the aim to find as many common characteristics in various experiments as possible. Moreover, at the same time, an attempt will be made to find a way to overcome the significant differences between them. The problem will be investigated with respect to data and measurement organization.

## 2.1 Draft look

Nuclear measurement generally aims at examining physical phenomena alias physical events by means of detectors. In the detectors physical events cause known responses which manifest themselves by electrical pulses. These pulses are digitized by converters, registered and stored by DAQ systems. On-line monitoring of the measurement has to ensure its error-free continuance. To obtain final physical results the measured data are usually subject to complex and time consuming processing procedures.

## 2.2 Data

### 2.2.1 *Event*

When a physical event occurs, it is registered by the detectors. Detector pulses are digitized by converters. These numerical values compose an *Event*. The *Event* is the digital counterpart of a physical event. The *Event* is composed of the data with a native converter structure and format. Usually the convertors of various types produce data with dissimilar format and structure. An *Event* parsing procedure is necessary if measured data have to be processed.

### 2.2.2 Converter *sub-event*

The converter is a hardware unit which can generally digitize pulses coming from more detectors at the same time. Then one can say that such a converter has more channels. On the other hand, all pulses caused by a physical event are as a rule digitized by more than one converter. Let us call the part of the *Event* generated by single converter as a converter *sub-event*. Then the complete *Event* can be composed of the one or more *converter sub-events*.

### 2.2.3 *Raw* **format**

Let us remind that the *converter sub-event* data structure differs between the various types of converters. Clearly the knowledge of every converter *sub-event* structure is necessary for the *Event* parsing. Since the various converters can produce differently formatted data, any changes in converter arrangement cause a need to recode the *Event* parsing procedure. So if the *Event* data would have a standardized format then the *Event* parsing procedure could be coded only once. Let us name such a format as a *raw-format*. This *raw-format* has to ensure that the converter *sub-event* data are unambiguously identifiable in the *Event*, the individual data items are unambiguously located there and their channel assignment is known. Let us postulate that the converter *sub-event* is composed of a *raw-header*, *pattern* and converted values. The *raw-header* identifies unambiguously the beginning of the *sub-event* in an *Event*. Moreover it has to contain the converter unique identifier and the number of the converted values. The *raw-pattern* determines the channel allocation. The word *raw* emphasizes the fact that this format is related to the native converter data. Usually the simpler converters produce their raw data in *raw-format* natively. Converter data with too complicated structure have to be parsed in a dedicated procedure.

### 2.2.4 *Module*

Now let us introduce an abstract counterpart of the hardware converter and call it a *module*. The hardware converter is identifiable by the *module* name and its unique address. The *module* has to contain the description of the data structure of its hardware counterpart in terms of the *raw-format*. It enables the *Event* parsing procedure to locate this *sub-event* in the data stream and to parse it. A reference to the dedicated parsing procedure has to be given if the converter data structure is too complex and cannot be described in terms of the *raw-format*. It will be helpful to introduce a so called *virtual module* extending the *module* role for description of arbitrary data acquired without hardware source.

### 2.2.5 *Raw-Event*

Let us introduce a *raw-Event* term. It is the name for a physical event counterpart composed of the corresponding converter *sub-events* whose data structures comply to the *raw-format*. Thanks to the uniform *raw-format* the *raw-Events* can be processed by a procedure which is not impacted by changes in the converter arrangement ( see fig.1. ).

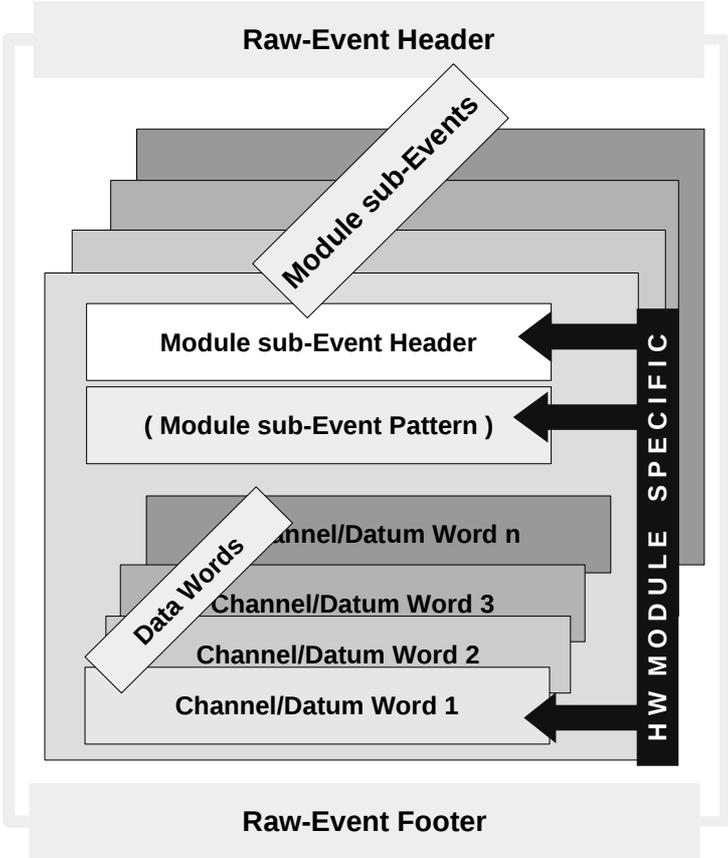

**Figure 1**. Structure of the *raw-Event*.

### 2.2.6 *Unified-Event*

The data in the *raw-Events* are identifiable by the converter identifiers and the channel numbers which correspond to the inner converter numbering. So the *raw-Event* can contain data related to different converters but with the same channel numbers. Moreover the converters rearrangement can change the converter identifiers and the physical meaning of the channels. So if an on-line monitoring and off-line processing would be based on the *raw-Events*, then a modification of the experimental setup would probably force us to recode the monitoring and processing procedures.

Let us renumber the native channels of all converters so that each datum value has assigned an additional unique channel identifier. It can be simply performed in the *Event* parsing procedure by means of a unique channel offset defined in the *module*. Then we can remove *raw-headers, raw-patterns* and *raw-footers* from the module *sub-Events* because *module* data are unambiguously identified only with their unique channel identifiers. Let us replace every *raw-Event* header and *raw-Event* footer with a new *unified-header* and a *unified-footer,* respectively. Both have to be unambiguously identifiable. So we have defined a *unified-Event*. It is a counterpart to physical event which is fully hardware independent ( see fig.2. ).

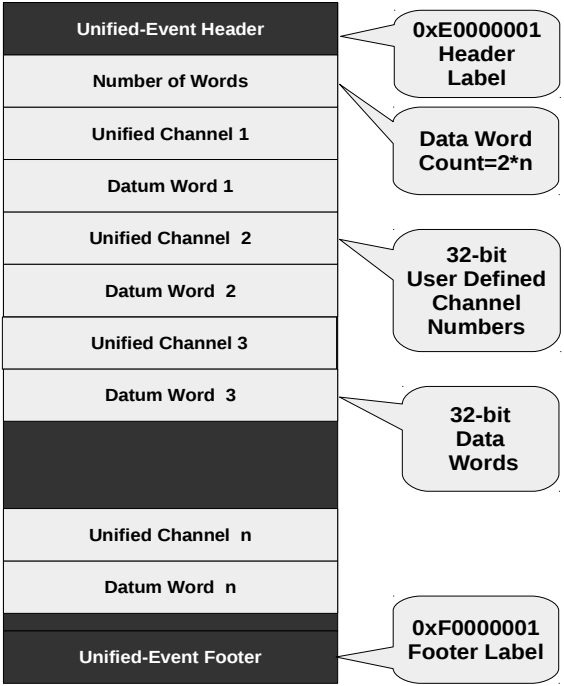

**Figure 2**. Structure of the *unified-Event*.2.3 Controlling and executing procedure

## *2.3 Controlling and executing procedure*

### 2.3.1 *Command*

The measurement hardware setup is designed according to the logic reflecting the physical aim of the experiment. Let us introduce a *command* as the elementary instruction for measurement control and execution. The *command* has its name, input parameters and instruction how to handle the result.

### 2.3.2 *Section*

Let us introduce a *section*. C*ommands* can be grouped in a *section*. A s*ection* is a container which allows to handle the *commands* inside en bloc. S*ections* can have names. Every name has to be unique among the other *section* names for unambiguous identification. The *section* can contain an instruction to determine how the *commands* inside it have to be handled. Moreover the *section* structure allows the use of *section commands* as branching points.

### 2.3.3 *Action*

The *action* ( see below ) is a container to associate several *sections* and *commands* together. It allows the use of *action commands* as branching points. The *actions*, *sections* and *commands* are  structural components from which the top level structure of the measurement procedure is built.

### 2.3.4 *Framework* and *framework segment*

Let us postulate that the top level structure of the measurement process can be composed of several phases. The change of the current phase is a consequence of the execution of one of the several special *orders*. Let us introduce a *framework* concept as the abstract counterpart of the top level measurement process structure. Then the *framework* will be composed of the so called *segments* which correspond the measurement phases ( see fig.3. ).

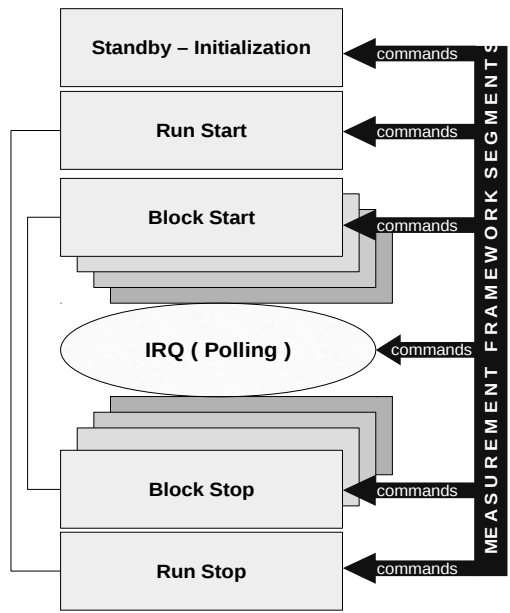

**Figure 3**. Structure of the measurement process.

The correspondence between the phases and the *segments* can be seen hereinafter:

| order | phase | framework segment |
|---|---|---|
| start measurement | initialization | *standby* |
| start run | set run conditions | *run_start* |
| start block | set block conditions | *block_start* |
| data taking | collect data | *interrupt, polling* |
| stop block | clean after block | *block_stop* |
| stop run | clean after run | *run_stop* |
| stop measurements | clean all | |

The phase arrangement involves several premises. The measurement setup has to be initialized before launching the data acquisition. The experimental data are gathered per batches into the files during the time intervals ( blocks ). The files are associated together and the group of files is denoted as a run ( see fig.4 ). Every run can be measured under different conditions. The data taking phase can be operated in two modes. The first, *interrupt* mode means that the DAQ system waits for an interrupt request to take pending data. The second, *polling* mode consists of periodic questioning whether some data are available.

The *framework segments* act as containers to arrange the *actions*, *sections* and *commands*.

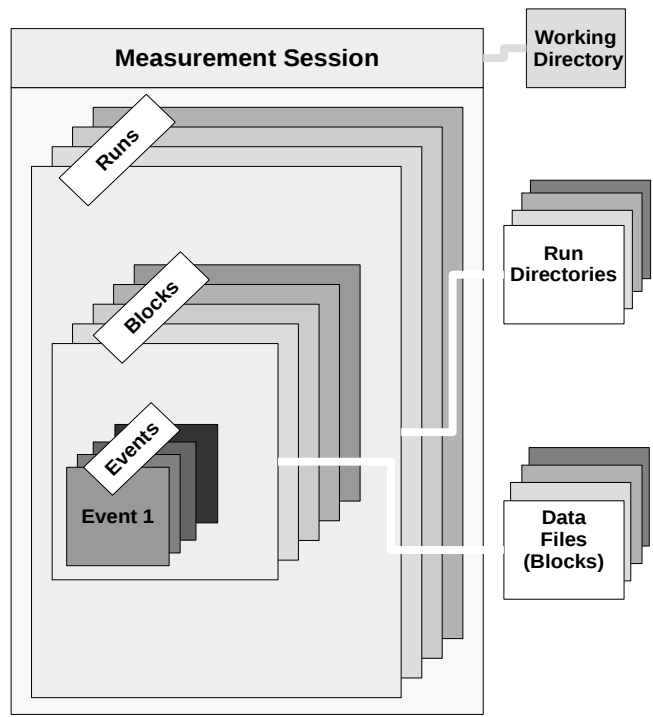

**Figure 4**. Data organization on a disc storage.

### 2.3.5 *Module* containers

To ensure the required measurement implementation, the *framework segments* have to be filled with s*ections* containing suitable *commands*. It is reasonable to group the *commands* relating to the particular hardware unit via corresponding *modules*. So let us introduce *module command* containers. The *module* can hold ( see the *framework segments* ) seven containers - *actions*: *standby*, *run_start*, *block_start*, *polling*, *interrup*t, *block_stop* and *run_stop*. The *action* (as defined above) is a contribution of the *module* to given *framework segment* ( see fig.*5* ).

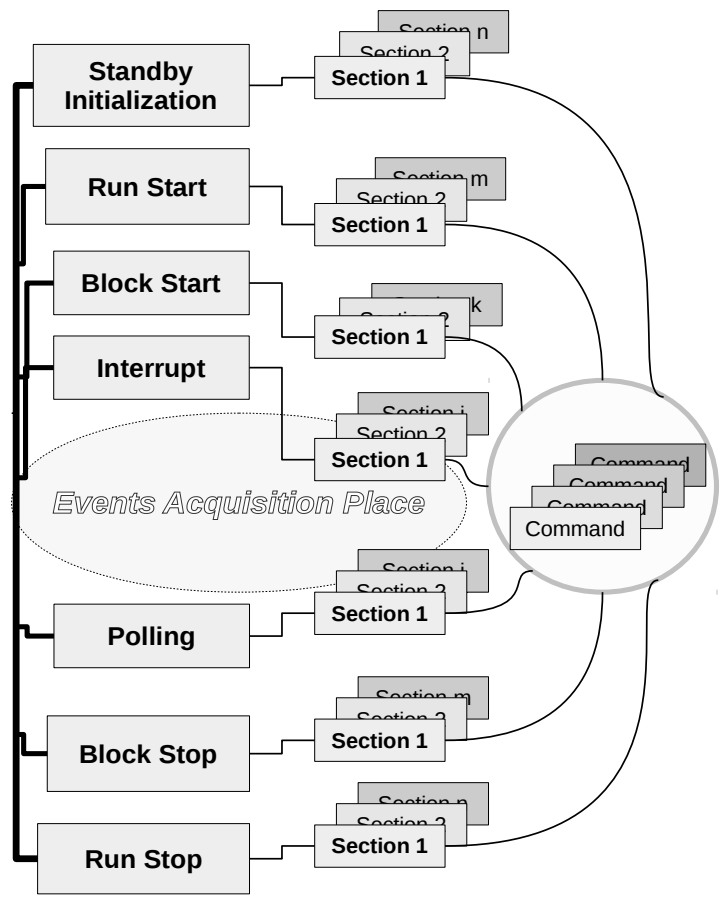

**Figure 5**. How DAQ commands are grouped and organized in module command containers.

The example of the VME CAEN V965 QDC *module* implementation can be seen hereinafter:

```
<module name="V965_2" base_GEO_address="0x100000"  crate="0"
base_24_address="" base_32_address="0x60020000" GEO_value="2">
    <channels unified_channel_offset="201" channel_count="32"/>
    <module_event>
        <header>
            <module_identifier GEO_address_mask="0xf8000000"
                        GEO_address_shift="27" crate_mask="0x00ff0000"
                        crate_shift="16"/>
            <identifier value="0x02000000" mask="0x07000000"/>
            <data_word_count mask="0x00003f00" shift="8"/>
        </header>
    <data data_word_count="HEADER_DEFINED" channel="YES">
            <module_identifier GEO_address_mask="0xf8000000"
                        GEO_address_shift="27" crate_mask="0x00ff0000"
                        crate_shift="16"/>
            <identifier value="0x00000000" mask="0x07000000"/>
            <not_valid value="0x06000000" mask="0x07000000"/>
            <value mask="0x00000FFF" shift="0"/>
            <channel mask="0x001f0000" shift="16"/>
        </data>
        <footer>
            <module_identifier GEO_address_mask="0xf8000000"
                        GEO_address_shift="27" crate_mask="0x00ff0000"
                        crate_shift="16"/>
            <identifier value="0x04000000" mask="0x07000000"/>
        </footer>
    </module_event>
    <actions>
        <standby>
            <section>
                <command name="CAENVME_WriteCycle"
                        offset_32_address="0x1006" address_modifier="0x09"
                        data_width="16" input_value="0x80" repeat="1"/>
                <command name="CAENVME_WriteCycle"
                        offset_32_address="0x1008" address_modifier="0x09"
                        data_width="16" input_value="0x80" repeat="1"/>
            </section>
        </standby>
        <run_start>
            <section>
                <command name="CAENVME_WriteCycle"
                        offset_32_address="0x1040" address_modifier="0x09"
                        data_width="16" input_value="0x0000" repeat="1"/>
                <command name="CAENVME_WriteCycle"
                        offset_32_address="0x100A" address_modifier="0x09"
                        data_width="16" input_value="0x0000" repeat="1"/>
            </section>
        </run_start>
        <run_stop>
            <section>
                <command name="CAENVME_WriteCycle"
                        offset_32_address="0x100A" address_modifier="0x09"
                        data_width="16" input_value="0x0000" repeat="1"/>
```

```xml
            </section>
        </run_stop>
        <block_start>
            <section>
                <command name="CAENVME_WriteCycle"
                        offset_32_address="0x1034" address_modifier="0x09"
                        data_width="16" input_value="0x0006" repeat="1"/>
            </section>
        </block_start>
        <block_stop>
            <section>
                <command name="CAENVME_WriteCycle"
                        offset_32_address="0x1032" address_modifier="0x09"
                        data_width="16" input_value="0x0006" repeat="1"/>
            </section>
        </block_stop>
        <interrupt>
            <section>
                <command name="CAENVME_ReadCycle"
                        offset_32_address="0x100E" address_modifier="0x09"
                        data_width="16" output_deposit="0"
                        result="RESTODEPOSIT"/>
                <command name="SPREAD8BITSTO07COMMONDEPOSITS"
                        input_deposit="0" input_svalue="DEPOSIT"/>
                <command name="TESTCOMMONDEPOSIT" common_deposit="0"
                        test="EQ" test_value="1" success="CONTINUE"
                        failure="BREAKSECTION"/>
                <command name="CAENVME_ReadCycle"
                        offset_32_address="0x0000" address_modifier="0x09"
                        data_width="32" result="SAVE"/>
                <command name="GETDATACOUNTPLUS1"
                        input_svalue="LAST_RESULT"
                        output_common_deposit="0" result="RESTOCDEPOSIT"/>
                <command name="CAENVME_BLTReadCycle"
                        offset_32_address="0x0000" address_modifier="0x0B"
                        data_width="32" how_smany_times="CDEPOSIT"
                        input_common_deposit="0"  result="SAVE"/>
            </section>
        </interrupt>
        <polling>
        </polling>
    </actions>
</module>
```

## 2.4 Summary

It appears the analysis performed above can serve as a resource for development of largely universal DAQ system. While the *framework segments* and the *raw-Event* to the *unified-Event* translator can be ultimately coded as simple cycles with a primitive logic, the diversity of the hardware arrangements and particular measuring logic can be described inside appropriate *modules* by means of the *commands* and at the same place the *raw-format* template can be used to describe converter sub-events for successful translation of their data into the *unified-format*. Then these *module* definitions can be loaded into the *framework segments* and the *Event* parser from a configuration file.

So finally we have a universal computer program and many various configuration files reflecting variety of measuring schemes and converter data. As it will be shown later the ability to translate data into the *unified-format* provides promising ground to find an algorithm for monitoring and data processing ( see fig. 6 ).

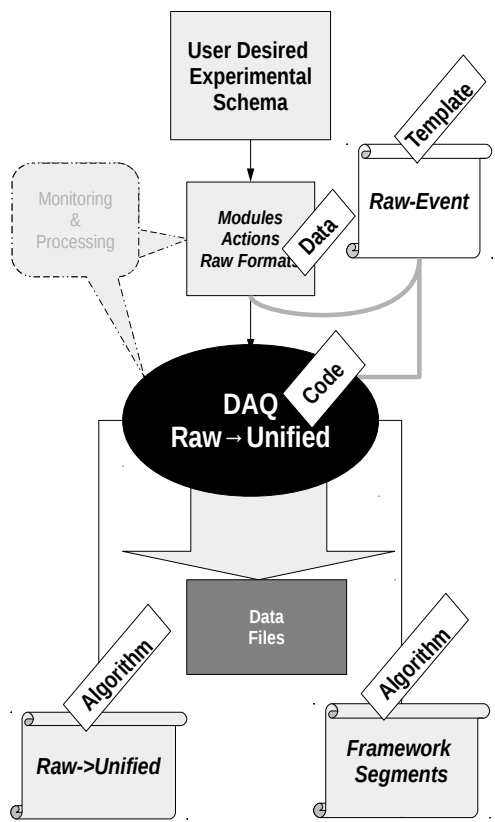

**Figure 6**. DAQ and *raw* to *unified* translation schema.

## 3. Abstract description of the monitoring and processing parts

Let us suppose that a histogram ( e.g. as understood in ROOT®[9] ) offers satisfactory presentation form of data processing results. Let us remember that the experimental data are stored via the *unified-Events* and that one *unified-Event* corresponds to the single physical event. Then the *unified-Event* is a native source for the histogram content. However, usually it is not enough to present only original converter data. So we have to provide a possibility to process the *unified-Event* data before they are added into histograms.

### 3.1 *Histogram*

Let us introduce an abstract object named *histogram* as monitoring and processing unit ( see fig. 7 ). It has to contain a histogram definition according to the ROOT requirements: the histogram's bin number and the low and upper limit of the first and last bin, say for example *nch*, *start_channel* and *stop_channel* parameters, respectively. Further, we need a list of *unified-Event* items which will contribute to the histogram.

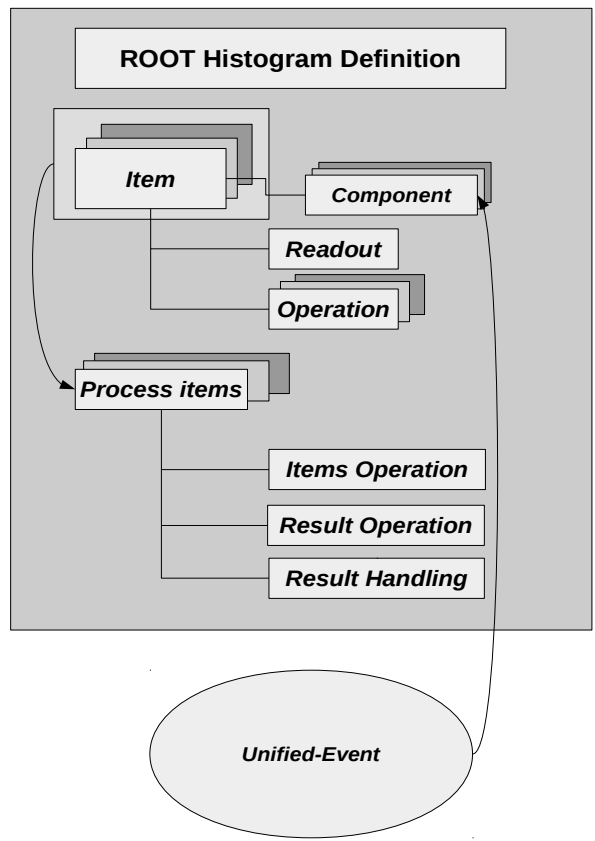

**Figure 7**. The histogram concept.

### 3.1.1 *Item* and *component*

Let us introduce a basic *histogram* building block called an *item*. A histogram will be built on the basis of a list of the *items*. The same *item* can be presented in the *item* list several times. In general, the *item* is an array of *components*. The *component* holds a [*value,channel,validity*] triplet where the [datum,channel] is as a rule a pair from a *unified-Event*. The low and upper limits of the *component* range are defined by *channel_start* and *channel_stop* parameters.

The *item* is characterized by its *type*. The *type* can have two values: *RELEVANT* or *CONDITION*. The *RELEVANT type* means that the *item* can potentially contribute to the histogram. The *CONDITION type* means that the *item* will only take part in logical operations.

#### 3.1.1.1 *Item readout*

The content of a *value* of each *component* of an *item* is defined by a *readout*. The *item values* can be loaded with a data value, *unified* channel value or logical value according to the value of the *item readout subject*: *DATA*, *CHANNEL* or *EXISTENCE*. The *component channel* always holds a *unified* channel number. The *validity* is a logical value, it has a logical meaning: the *component* is valid or not. So this value can be true or false. The *item values* and *validity* values can be changed during the *item* processing.

#### 3.1.1.2 *Item operation*

The *item values* can be modified by means of arithmetic operations and can take part in logical operations. The arithmetic operations change the *item values* while the logical operations use the *item validity* value. The *item operation* is defined by means of an *operation type* and its *parameters*. It is possible to execute several operations on the same *item*. If the *item* is composed of more than one *component* the operations are executed for each *component* separately.

### 3.1.2 *Process_items item* group

The *items* can be associated into a *process_items* group. The same *item* can be simultaneously assigned to different groups. Introducing the groups gives the user the possibility to perform arithmetic and logic operations among the grouped *items*. The groups are evaluated step by step. The *item* association into the groups also gives possibility to apply subsequently various operations to the results obtained from the previous group evaluations.

#### 3.1.2.1 *Result_operation*

The final result can be modified by a *result_operation*. It can be an arithmetic operation affecting the result value itself or a logical one which determines whether the result value is valid or not.

### 3.1.2.2 *Handling_method*

If the result is valid, a *handling_method* defines the way how the histogram will be updated. If the *handling_method* is *PRELIMINARY,* the histogram is not immediately updated. This is used when a *histogram* requires more complex processing than is possible to define inside only one *process_items* group.

The example of a *histogram* implementation follows:

```
<histogram scope_x="EVENT" start_channel_x="0" nch_x="4096"
                          name="channel_5_6-T6-E+dE">
      <item_x type="RELEVANT" channel_start="5" channel_stop="5">
            <readout source="EXTERNAL" subject="DATA"/>
            <operation type="RANDOMIZE"/>
            <operation type="TRANSFORM" operation="POLYNOM"
                          a0="0." a1="1.0"/>
      </item_x>
      <item_x type="RELEVANT" channel_start="6" channel_stop="6">
            <readout source="EXTERNAL" subject="DATA"/>
            <operation type="RANDOMIZE"/>
            <operation type="TRANSFORM" operation="POLYNOM"
                          a0="0." a1="1.0"/>
      </item_x>
      <process_items_x>
            <item_x index="0"/>
            <item_x index="1"/>
            <items_operation type="ARITHMETIC" operation="SUM"/>
            <result_operation type="iTRANSFORM" operation="POLYNOM"
                          a0="0." a1="1.0"/>
            <handling method="INCREMENT" result="ADDRESS"
                          antialiasing="YES"/>
      </process_items_x>
</histogram>
```

### 3.2 Summary

The *unified* format and the *histogram* concept ( see fig. 8 ) serve as a basis for successful algorithmization of data monitoring and processing. The *item operations*, *process_items operations*, *result_operation* and *handling_method* serve as containers which can be filled with commands by means of a simple processing language. It allows many ways to process data before they are presented via histograms. The data processing and visualization can be easily coded as a sequence of simple instructions and, if necessary, can be easily updated. User's requirements can be easily loaded as a configuration file.

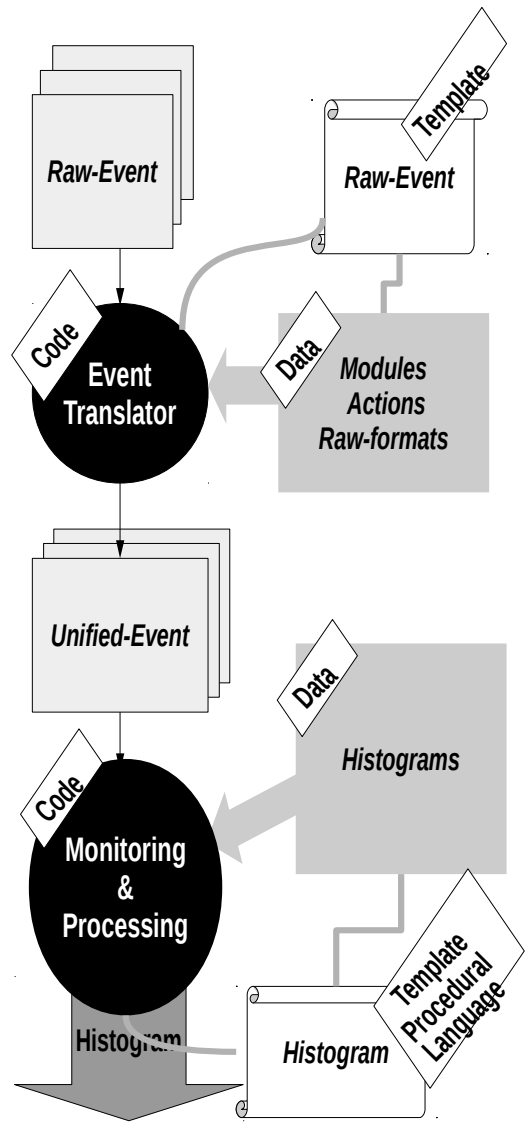

**Figure 8**. Monitoring and data processing schema.

## 4. System architecture

The analysis and the resulting design presented so far do not leave a conceptional and theoretical level. However, the realization of the presented ideas in the practical nuclear experiments means implementing them in an environment of particular technologies, instruments and standards.

All more complex modular and scalable measuring systems show characteristic similarities. They are composed of hardware and software parts. The hardware is built around a standardized bus. The software usually runs on a personal computer with an operating system. The OS offers basic instruments for exploiting the hardware and software system resources. These circumstances determine the architecture and properties of every system and establish the obligatory environment for development of applications.

Our environment for development was: PC computer, Linux®[10] operating system, CAMAC and VME standards, networking, C and C++ programming languages, XML instruments, Qt®[11] graphic library, CERN's ROOT[9] – graphic analyzing framework. The system was devised as two collaborating multi-layer parts. First, the data taking system and, second, the system for data monitoring, processing and administration. They communicate with each other using TCP/IP. The typical hardware arrangement is shown on fig. 9, while the software architecture on fig. 10.

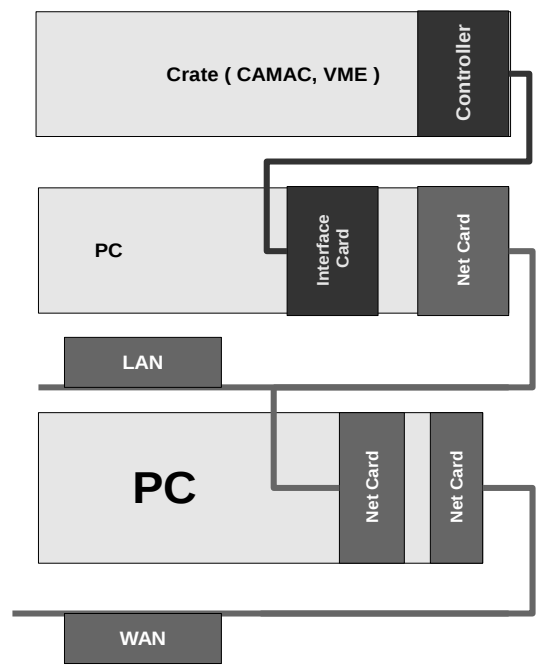

**Figure 9**. Typical hardware arrangement of a simple measurement setup.

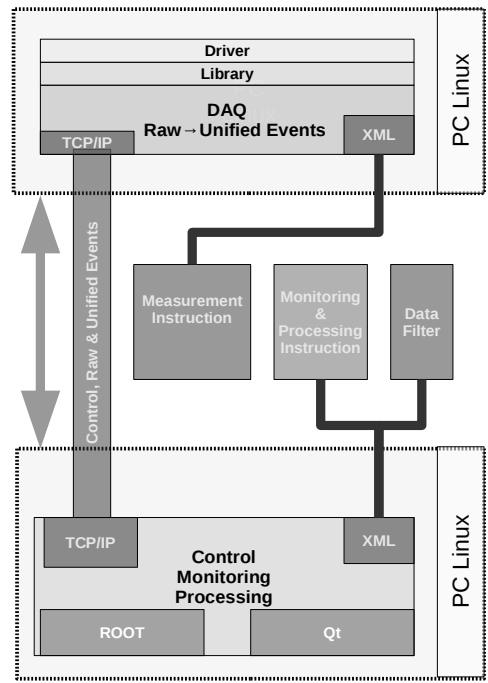

**Figure 10**. Software architecture of the whole system.

## *4.1 DAQ design*

### 4.1.1 Device driver

An operating system driver forms the lowest DAQ layer. It allows to control the crate bus and the connected hardware modules by means of operating system tools. It is obvious that different buses ( CAMAC or VME ) standards require dedicated drivers. In adition, different crate controllers need as a rule different drivers in spite of the fact that they control the same bus. The driver ensures that the hardware resources connected via the crate bus are visible as the operating system standard device. Such an approach has appreciable advantages: we can use standard, reliable and effective tools of the operating system. Let us mention the interrupt system as an example. However, it has also an unfortunate disadvantage: system layers are inserted between the hardware and the application software  slowing down the execution.

### 4.1.2 Command library

Accessing the hardware resources via the driver enables us to use them fully. However, the communication format is far-away from the desirable user-friendly format of the *commands*.  Let us introduce another layer formalized as a library. Its functions will encapsulate all driver functions in the format close to the user-friendly *commands*. Both the driver layer and library are tied to the hardware standard which was used for the experimental setup. If this standard is changed, both layers have to be rewritten or replaced. On the other hand, the driver and the *command* library separate the variety of the hardware standards from the higher layer. Thus if the hardware standard is changed the forced changes in the higher application layer are significantly reduced, if any at all.

## 4.1.3 DAQ layer

The third layer implements the ideas described above. It was projected as a formation of several collaborating threads. These are the main administration thread providing the communication with the outer world, the DAQ thread, the thread of the *raw->unified* data transformation, the pair of threads storing the *raw* and *unified* data onto the recording device and the pair of threads where two network servers are running and making the *raw* and *unified* data available.

### 4.1.3.1 Administration thread

The main administration thread supports two kinds of communication: local and remote. This  communication thread receives control instructions and decodes and executes them. It makes possible debugging of the DAQ system because it can work as local or remote console. The configuration file describing the *modules*, *virtual modules* and the content of the *framework segments* is loaded, decoded and executed here.

### 4.1.3.2 DAQ thread

The DAQ thread is created by the main thread after correct system and hardware initialization and after the *standby segment commands* are executed. Then it is waiting for transfer to the start run phase. It takes data in two basic modes. Either waits for an interrupt and then performs the *interrupt segment commands* or periodically executes the *polling segment commands*. Moreover, the DAQ thread also guards the size or duration of data block acquisition and run scheduling.

### 4.1.3.3 Translator thread

The next thread serves as a translator of the data in the *raw* format to the *unified* one. It parses the *raw* buffer and stores the translated data to the *unified* buffer. During the transformation, the thread checks *raw* data structure and logs ones which do not correspond to any data template declared in the *modules*. The thread can be optionally switched off and the *raw* to *unified* transformation can be completed off-line later.

### 4.1.3.4 Threads to store data

Storage of data is performed by a pair of threads. The first thread stores the *raw* data and the second one stores the *unified* data onto the recording device. They can be switched on or off independently. Their activity is triggered by the DAQ thread. The *unified* data storing thread can be switched on only when the *raw* to *unified* translator thread is running.

### 4.1.3.5 Communication threads

Finally the network server threads can run. They work independently and send the *raw* and *unified* data to connecting clients. The dispatching frequency is configurable. The *unified* server can be started only when the *raw* to *unified* transformation thread is running.

## 4.2 Administration, monitoring and processing design

This software layer acts as a controller of the DAQ system and provides both on-line monitoring and off-line processing. It involves data visualization, sorting and processing and preparation of the data for more advanced off-line analysis. It communicates with the DAQ system using a TCP/IP connection. Both monitoring and processing does not depend on the origin of the data and can be used for processing of any data which respect the *unified* format. It uses Linux®[10] operating system and Qt®[11] and ROOT®[9] environment for user interface and for the data visualization and analysis. It consists of the three layers. Each of them uses its own window for user interface.

## 4.2.1 Administration layer

The administration layer utilizes the server – client technology for the communication with the DAQ system. The server side is implemented in the DAQ system. This connection is used for sending control commands and receiving status and error information. The network connection is negotiated and utilized by means of a console. The administration module supports up to 8 consoles at the same time to control multiple DAQ either independently or synchronously. Besides a graphic interface the console also provides the DAQ command prompt. Furthermore, the administration module allows to start and stop the measurement process and to force the DAQ system to re-initialize itself by loading another configuration file. If the DAQ system waits in the *standby* phase the user can apply the whole set of the *commands* by means of the command prompt. This mode is very useful especially for debugging purposes. If the file system of the DAQ host is connected to the administration PC file system ( for instance NFS ) the remote regime of the data processing is available.

## 4.2.2 Monitoring and processing layers

Another server – client channel has been added to ensure connection between the DAQ system and the monitoring and processing layers. Two sockets are utilized: the first socket for sending the *raw-Events*, the second one for *unified-Events*. The monitoring layer can visualize the data either on-line via the network connection or off-line from the files previously stored on the PC. To work with the remote data in the off-line mode, NFS is required. The *raw* data can be presented only in text mode while the *unified* data can be presented both in text and graphically as histograms. In on-line regime the histograms are continually updated as the *unified-Events* coming one by one.

The structure of the histogram tree and the manner of the data visualization are defined in the configuration file by the *histogram* objects. The configuration file can be loaded at any time. It means that the style of the data presentation can be changed during the data taking. The data monitoring can be stopped and started repeatedly. The histograms can be stored on the disk, printed and selectively cleared.

The processing layer, compared to the monitoring one, adds the possibility to store the histograms in a simple format ( ASCII and binary ) suitable for a subsequent proprietary processing. The histogram definition for both the processing and monitoring is one and the same. Thus if the advanced off-line data processing is not too time consuming it can be also used for on-line data visualization.

The data, which the off-line processing is related to, can be chosen according to many conditions. The selection criteria can be date and time, run and block numbers and user labels which are part of the data file names and directory names created during the data taking.

Besides the possibilities integrated in the ROOT® system, the monitoring and processing layers contain custom tools for processing one and two dimensional spectra. They include determining the peak position and area. The peak shape can be approximated with several analytical shapes and energy and FWHM calibration is supported. Additionally, it is possible to process two dimensional matrices obtained in coincidence measurements.

## 5. Practical implementation

The ideas presented above were realized as real CAMAC and VME DAQ systems. The system is implemented as a set of several programs: `nwcamac` – CAMAC DAQ program, `nwvme` – VME DAQ program, `nwgo` – graphic interface to control `nwcamac` and `nwvme`. The monitoring and processing programs for CAMAC and VME DAQ programs are the same: `namon` and `nwview`, which is a lightened version of `namon` working only with on-line data.

The CAMAC DAQ systems are built on the author's driver and command library tied with the KK009[16] CAMAC controller developed in JINR Dubna. The VME DAQ systems use the CAEN[15] Linux driver and CAENVMELibs.

The monitoring and processing parts do not have to be used only with a native DAQ systems. Its usefulness can be much wider. In several cases it is(was) utilized in cooperation with as different systems as is the GVD-Baikal[13] detector or the DIRAC[3] DAQ system.

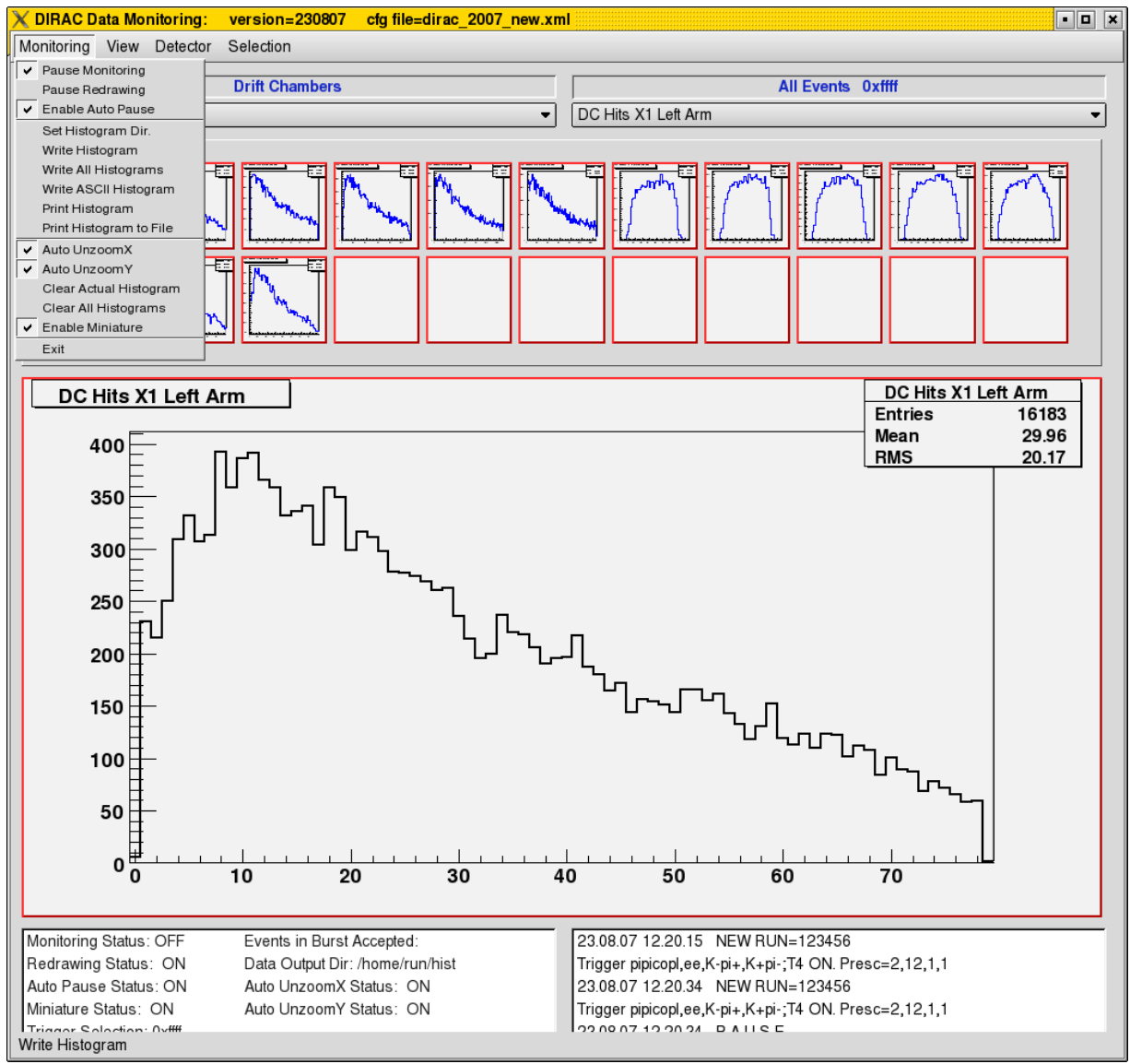

**Figure 11**. Typical monitoring window – DIRAC project.

On the whole the described system has been applied and is used in several experiments in Joint Institute for Nuclear Research (Dubna)[1,4,5,12,13], Nuclear Physics Institute ( Czech republic )[2] and CERN[3]. These experimental projects have quite different data formats, experimental setups and physical aims. Nevertheless, this variety can be hidden behind appropriate configuration files whereas the DAQ, monitoring and processing programs are the same. There is the only exception in the case of the DIRAC experiment in CERN. The data being monitored there do not come via network but they are available by means of a shared memory pool. Typical monitoring window ( DIRAC ) is shown on fig.11.

## 6. Example

Let us suppose that we are going to measure gamma spectrum of some radioactive isotope. Let us assemble very simple spectrometer which consists of:

a CAEN VME crate, the VME-PCI Bridge (V2718) + PCI OpticalLink (A2818) and the V785N 16 channel peak sensing ADC placed in position number 5 of the crate. Gates will be prepared outside of the VME crate. The measurement will be organized event by event. We would like to mark every event with real time stamp. Pulses from spectroscopic amplifier will be put in the ADC input number 0. Data will be stored in files every 5 minutes. The whole measurements will last 30 minutes.

The DAQ configuration file can look as follows:

```xml
<?xml version = '1.0'?>
<nwvme_cfg>
<!-- virtual module generating sub-event containing system time stamp →
    <module name="LINUX_TIME_1" crate="0" GEO_value="33" bit16_used="YES">
        <channels  channel_count="4" unified_channel_offset="1024"/>
        <module_event>
                <header value="0xfa000421">
                        <identifier value="0x02000000" mask="0x07000000"/>
                 <data_word_count mask="0x0F00" shift="8"/>
            </header>
                <pattern value="0x000f"/>
            <data data_word_count="HEADER_DEFINED">
                <value mask="0xFFFF"/>
            </data>
                    <footer value="0xfc000000">
                        <identifier value="0x04000000" mask="0x07000000"/>
                    </footer>
        </module_event>
    </actions>
        <standby>
        </standby>
        <run_start>
          <section>
            <command name="INSERTVALUE" input_svalue="CRATE_HEADER" result="SAVE"/>
            <command name="INSERTVALUE" input_svalue="MODULE_HEADER" result="SAVE"/>
            <command name="INSERTVALUE" input_svalue="MODULE_PATTERN" result="SAVE"/>
            <command name="GETLINUXTIME" result="SAVE"/>
            <command name="INSERTVALUE" input_svalue="MODULE_FOOTER" result="SAVE"/>
            <command name="INSERTVALUE" input_svalue="CRATE_FOOTER" result="SAVE"/>
          </section>
        </run_start>
        <run_stop>
          <section>
            <command name="INSERTVALUE" input_svalue="CRATE_HEADER" result="SAVE"/>
            <command name="INSERTVALUE" input_svalue="MODULE_HEADER" result="SAVE"/>
            <command name="INSERTVALUE" input_svalue="MODULE_PATTERN" result="SAVE"/>
            <command name="GETLINUXTIME" result="SAVE"/>
            <command name="INSERTVALUE" input_svalue="MODULE_FOOTER" result="SAVE"/>
            <command name="INSERTVALUE" input_svalue="CRATE_FOOTER" result="SAVE"/>
          </section>
        </run_stop>
        <block_start>
          <section>
            <command name="INSERTVALUE" input_svalue="CRATE_HEADER" result="SAVE"/>
            <command name="INSERTVALUE" input_svalue="MODULE_HEADER" result="SAVE"/>
            <command name="INSERTVALUE" input_svalue="MODULE_PATTERN" result="SAVE"/>
            <command name="GETLINUXTIME" result="SAVE"/>
            <command name="INSERTVALUE" input_svalue="MODULE_FOOTER" result="SAVE"/>
            <command name="INSERTVALUE" input_svalue="CRATE_FOOTER" result="SAVE"/>
          </section>
        </block_start>
        <block_stop>
```

```xml
            <section>
                <command name="INSERTVALUE" input_svalue="CRATE_HEADER" result="SAVE"/>
                <command name="INSERTVALUE" input_svalue="MODULE_HEADER" result="SAVE"/>
                <command name="INSERTVALUE" input_svalue="MODULE_PATTERN" result="SAVE"/>
                <command name="GETLINUXTIME" result="SAVE"/>
                <command name="INSERTVALUE" input_svalue="MODULE_FOOTER" result="SAVE"/>
                <command name="INSERTVALUE" input_svalue="CRATE_FOOTER" result="SAVE"/>
            </section>
        </block_stop>
        <interrupt>
            <section>
                <command name="INSERTVALUE" input_svalue="MODULE_HEADER" result="SAVE"/>
                <command name="INSERTVALUE" input_svalue="MODULE_PATTERN" result="SAVE"/>
                <command name="GETLINUXTIME" result="SAVE"/>
                <command name="INSERTVALUE" input_svalue="MODULE_FOOTER" result="SAVE"/>
            </section>
        </interrupt>
        <polling>
            <section polling_divider="1">
                <command name="INSERTVALUE" input_svalue="MODULE_HEADER" result="SAVE"/>
                <command name="INSERTVALUE" input_svalue="MODULE_PATTERN" result="SAVE"/>
                <command name="GETLINUXTIME" result="SAVE"/>
                <command name="INSERTVALUE" input_svalue="MODULE_FOOTER" result="SAVE"/>
            </section>
        </polling>
    </actions>
</module>

  <module name="V785" base_GEO_address="0x280000"  crate="0" base_32_address="0x10050000"
                                                                       GEO_value="5">
  <channels unified_channel_offset="1" channel_count="16"/>
  <module_event>
    <header>
        <module_identifier GEO_address_mask="0xf8000000" GEO_address_shift="27"
                                          crate_mask="0x00ff0000" crate_shift="16"/>
        <identifier value="0x02000000" mask="0x07000000"/>
        <data_word_count mask="0x00003f00" shift="8"/>
    </header>
    <data data_word_count="HEADER_DEFINED" channel="YES">
        <module_identifier GEO_address_mask="0xf8000000" GEO_address_shift="27"
                                          crate_mask="0x00ff0000" crate_shift="16"/>
        <identifier value="0x00000000" mask="0x07000000"/>
        <not_valid value="0x06000000" mask="0x07000000"/>
        <value mask="0x00000FFF" shift="0"/>
        <channel mask="0x001e0000" shift="17"/>
    </data>
    <footer>
        <module_identifier GEO_address_mask="0xf8000000" GEO_address_shift="27"
                                          crate_mask="0x00ff0000" crate_shift="16"/>
        <identifier value="0x04000000" mask="0x07000000"/>
    </footer>
  </module_event>
    <actions>
        <standby>
            <section>
                <!-- single shot reset -->
                <command name="CAENVME_WriteCycle" offset_32_address="0x1016"
                    address_modifier="0x09" data_width="16" input_value="0x0000" repeat="1"/>
                <!-- reset on -->
                <command name="CAENVME_WriteCycle" offset_32_address="0x1006"
                    address_modifier="0x09" data_width="16" input_value="0x80" repeat="1"/>
                <!-- reset off -->
                <command name="CAENVME_WriteCycle" offset_32_address="0x1008"
                    address_modifier="0x09" data_width="16" input_value="0x80" repeat="1"/>
                <!-- inhibit on, clear data, pointers and counter -->
                <command name="CAENVME_WriteCycle" offset_32_address="0x1032"
                    address_modifier="0x09" data_width="16" input_value="0x0106" repeat="1"/>
                <!-- no clear data, pointers and counter -->
                <command name="CAENVME_WriteCycle" offset_32_address="0x1034"
                    address_modifier="0x09" data_width="16" input_value="0x0004" repeat="1"/>
```

```xml
            <!-- zapis GEO -->
            <command name="CAENVME_WriteCycle" offset_32_address="0x1002"
                address_modifier="0x09" data_width="16" input_value="0x05" repeat="1"/>
            <!-- interrupt vector = geo -->
            <command name="CAENVME_WriteCycle" offset_32_address="0x100c"
                address_modifier="0x09" data_width="16" input_value="0x0005" repeat="1"/>
            <!-- event trigger register 1 event generates interrupt -->
            <command name="CAENVME_WriteCycle" offset_32_address="0x1020"
                address_modifier="0x09" data_width="16" input_value="0x0001" repeat="1"/>
            <!-- end of block bit -->
          <command name="CAENVME_WriteCycle" offset_32_address="0x1010"
                address_modifier="0x09" data_width="16" input_value="0x0004" repeat="1"/>
            <!-- thresholds default no thresholds, 0x0100 - KILL the channel-->
            <command name="CAENVME_WriteCycle" offset_32_address="0x1080"
                address_modifier="0x09" data_width="16" input_value="0x0048" repeat="1"/>
            <command name="CAENVME_WriteCycle" offset_32_address="0x1084"
                address_modifier="0x09" data_width="16" input_value="0x0100" repeat="1"/>
            <command name="CAENVME_WriteCycle" offset_32_address="0x1088"
                address_modifier="0x09" data_width="16" input_value="0x0100" repeat="1"/>
            <command name="CAENVME_WriteCycle" offset_32_address="0x108C"
                address_modifier="0x09" data_width="16" input_value="0x0100" repeat="1"/>
            <command name="CAENVME_WriteCycle" offset_32_address="0x1090"
                address_modifier="0x09" data_width="16" input_value="0x0100" repeat="1"/>
            <command name="CAENVME_WriteCycle" offset_32_address="0x1094"
                address_modifier="0x09" data_width="16" input_value="0x0100" repeat="1"/>
            <command name="CAENVME_WriteCycle" offset_32_address="0x1098"
                address_modifier="0x09" data_width="16" input_value="0x0100" repeat="1"/>
            <command name="CAENVME_WriteCycle" offset_32_address="0x109C"
                address_modifier="0x09" data_width="16" input_value="0x0100" repeat="1"/>
            <command name="CAENVME_WriteCycle" offset_32_address="0x10A0"
                address_modifier="0x09" data_width="16" input_value="0x0100" repeat="1"/>
            <command name="CAENVME_WriteCycle" offset_32_address="0x10A4"
                address_modifier="0x09" data_width="16" input_value="0x0100" repeat="1"/>
            <command name="CAENVME_WriteCycle" offset_32_address="0x10A8"
                address_modifier="0x09" data_width="16" input_value="0x0100" repeat="1"/>
            <command name="CAENVME_WriteCycle" offset_32_address="0x10AC"
                address_modifier="0x09" data_width="16" input_value="0x0100" repeat="1"/>
            <command name="CAENVME_WriteCycle" offset_32_address="0x10B0"
                address_modifier="0x09" data_width="16" input_value="0x0100" repeat="1"/>
            <command name="CAENVME_WriteCycle" offset_32_address="0x10B4"
                address_modifier="0x09" data_width="16" input_value="0x0100" repeat="1"/>
            <command name="CAENVME_WriteCycle" offset_32_address="0x10B8"
                address_modifier="0x09" data_width="16" input_value="0x0100" repeat="1"/>
            <command name="CAENVME_WriteCycle" offset_32_address="0x10BC"
                address_modifier="0x09" data_width="16" input_value="0x0100" repeat="1"/>

        </section>
    </standby>
<run_start>
    <section>
        <!-- clear event counter register -->
        <command name="CAENVME_WriteCycle" offset_32_address="0x1040"
            address_modifier="0x09" data_width="16" input_value="0x0000" repeat="1"/>
        <!-- interrupt level=2 0 means interrupt forbidden -->
        <command name="CAENVME_WriteCycle" offset_32_address="0x100A"
            address_modifier="0x09" data_width="16" input_value="0x0002" repeat="1"/>
    </section>
</run_start>
<run_stop>
    <section>
        <!-- interrupt level=0   interrupt forbidden -->
        <command name="CAENVME_WriteCycle" offset_32_address="0x100A"
            address_modifier="0x09" data_width="16" input_value="0x0000" repeat="1"/>
        <!-- single shot reset -->
        <command name="CAENVME_WriteCycle" offset_32_address="0x1016"
            address_modifier="0x09" data_width="16" input_value="0x0000" repeat="1"/>
    </section>
</run_stop>
<block_start>
    <section>
```

```xml
                    <!-- inhibit off -->
                    <command name="CAENVME_WriteCycle" offset_32_address="0x1034"
                            address_modifier="0x09" data_width="16" input_value="0x0006" repeat="1"/>
                </section>
            </block_start>
            <block_stop>
                <section>
                    <!-- inhibit on, clear data, pointers and counter -->
                    <command name="CAENVME_WriteCycle" offset_32_address="0x1032"
                            address_modifier="0x09" data_width="16" input_value="0x0006" repeat="1"/>
                </section>
            </block_stop>
            <interrupt>
                <section>
                    <command name="CAENVME_ReadCycle" offset_32_address="0x100E"
                        address_modifier="0x09" data_width="16" output_deposit="0"
                                                                result="RESTODEPOSIT"/>
                    <command name="SPREAD8BITSTO07COMMONDEPOSITS" input_deposit="0"
                                                                input_svalue="DEPOSIT"/>
                    <command name="TESTCOMMONDEPOSIT" common_deposit="0" test="EQ" test_value="1"
                                                success="CONTINUE" failure="BREAKSECTION"/>
                    <command name="INSERTVALUE" input_svalue="CRATE_HEADER" result="SAVE"/>
                    <!-- raw sub-event header  -->
                    <command name="CAENVME_ReadCycle" offset_32_address="0x0000"
                                        address_modifier="0x09" data_width="32" result="SAVE"/>
                    <command name="GETDATACOUNTPLUS1" input_svalue="LAST_RESULT"
                                        output_common_deposit="0" result="RESTOCDEPOSIT"/>
                    <command name="CAENVME_BLTReadCycle" offset_32_address="0x0000"
                        address_modifier="0x0B" data_width="32" how_smany_times="CDEPOSIT"
                                                input_common_deposit="0"  result="SAVE"/>
                    <command name="INSERTVALUE" input_svalue="CRATE_FOOTER" result="SAVE"/>
                </section>
            </interrupt>
            <polling>
            </polling>
        </actions>
</module>

<crate_event>
    <!-- definition of the raw-Event identificators -->
    <header value="0xB0000000"/>
    <footer value="0xB8000000"/>
    <module_header_GEO_ID mask= "0xf8000000" shift="27"/>
    <module_header_ID value="0x02000000" mask="0x07000000"/>
    <module_footer_ID     value="0x04000000" mask="0x07000000"/>
</crate_event>

<crate number="0">
        <interrupt mask="0x7e" timeout="400"/>
</crate>
  <framework>
    <standby>
        <section>
            <command name="CAENVME_SystemReset" />
        </section>
        <include module="V785" GEO_vsn="5"/>
    </standby>
    <run_start>
        <section>
            <command name="INSERTVALUE" input_svalue="RUN_HEADER" result="SAVE"/>
        </section>
        <include module="V785" GEO_vsn="5"/>
    </run_start>
    <run_stop>
        <include module="V785" GEO_vsn="5"/>
        <section>
            <command name="INSERTVALUE" input_svalue="RUN_FOOTER" result="SAVE"/>
        </section>
    </run_stop>
    <block_start>
```

```
        <include module="V785" GEO_vsn="5"/>
    </block_start>
    <block_stop>
        <include module="V785" GEO_vsn="5"/>
    </block_stop>
    <interrupt>
        <include module="LINUX_TIME_1" GEO_vsn="33"/>
        <include module="V785" GEO_vsn="5"/>
        <include module="LINUX_TIME_1" GEO_vsn="33"/>
    </interrupt>
    <polling>
    </polling>
  </framework>
  <run name="V785Nirq" file_extension="vme" number="0" wdir="/data/152Eu">
      <unified_thread switch_on="YES" recording_step="20000"/><!-- [us] -->
      <raw_data file="YES" recording_step="21000"/>
      <unified_data file="YES" recording_step="22000"/>
      <raw_stream switch_on="YES" service="9301" stream_step="100000"/>
      <unified_stream switch_on="YES" service="9302" stream_step="10000"/>
      <verbose level="0xc001"/>
      <header value="0xC0000000"/>
      <footer value="0xC8000000"/>
      <block size="12000000" quantity="6" duration="300000" number="0"/>
  </run>
</nwvme_cfg>
```

The configuration information for monitoring and data processing can look this way:

```
<?xml version = '1.0'?>
    <namon_data_inspection_cfg>
        <pool refresh_time="5000" header="0xE0000000" footer="0xF0000000"/>
        <detector name="Single Spectra">
            <histogram_group name="V785 Single_Spectra">
                <histogram scope_x="EVENT" start_channel_x="0" nch_x="4096" name="INPUT-0"
                                                                            process="YES">
                    <item_x type="RELEVANT" channel_start="1" channel_stop="1">
                        <readout source="EXTERNAL" subject="DATA"/>
                        <operation type="BLANK"/>
                    </item_x>
                    <process_items_x>
                        <item_x index="0"/>
                        <items_operation type="BLANK"/>
                        <result_operation type="BLANK"/>
                        <handling method="INCREMENT" result="ADDRESS"/>
                    </process_items_x>
                </histogram>
            </histogram_group>
        </detector>
    </namon_data_inspection_cfg>
```

# 7. Conclusion

An abstract concept of the nuclear physics measurement was presented. It was demonstrated that the suitable abstraction can lead towards the formal description usable for simple practical implementation of the universal DAQ, monitoring and processing systems. Practical usage of this system consists of two simple steps. First, the system has to be installed on the PCs. Second, the measurement setup, the data formats and the data processing requirements have to be described by means of simple pseudo languages and presented as XML configuration files.

## Acknowledgements

The work has been performed in the course of 2003 – 2010 years in JINR, NPI and CERN and financed by the 03-02-17395 RFFI, LA-INGO LA08015, ME643 MŠMT and ME902 MŠMT grants. The author is grateful to  V.Stegajlov, P.Čaloun, V.Olshevsky, K.Gritsay, V.Kroha, Š.Piskoř, V.Burjan and J.Mrázek for numerous fruitful discussions and to other collaborators for  their continual support.

## References

[1] V.G.Kalinnikov et al., Nucl. Instr. Meth. B 70 (1992) 62-68.
[2] V. Kroha et al., Proceedings of the Carpathian Summer School of Physics 2007. AIP
    Conference Proceedings, 972 (2008) 279-285.
[3] B. Adeva et al., Nucl. Instr. Meth. A 515 (2003) 467-496.
[4] Z.Hons, 54 International Meeting on Nuclear Spectroscopy and Nuclear Structure-2004
    Belgorod. (2004) 289-290.
[5] V.G.Kalinnikov, Z.Hons, V.I.Stegajlov, P.Caloun, 54.International Meeting on Nuclear
    Spectroscopy and Nuclear Structure-2004 Belgorod. (2004) 272-273.
[6] www.aps.anl.gov/epics/
[7] www.comedi.org
[8] www.ni.com/labview/
[9] root.cern.ch
[10] www.linux.com
[11] www.qt.io
[12] I.Alekseev, ..., Z.Hons,... arXiv:1305.3350.
[13] A.D. Avrorin,..., Z. Hons,... arXiv:1308.1833.
[14] A.Kh. Inoyatova,..., Z. Hons. J. Journal of Electron Spectroscopy and Related
    Phenomena 187 (2013) 61–64.
[15] www.caen.it/csite/Product.jsp?parent=11
[16] Georgiev, A., Churin, I.N. JINR-P10-88-381